\documentclass[prl,twocolumn,superscriptaddress,showpacs]{revtex4}
\usepackage{graphicx,epsfig}
\begin{document}
\title{Localized Perturbations of Integrable Systems}

\author{Saar Rahav}
\affiliation{Department of Physics, Technion, Haifa 32000, Israel.}
\author{Shmuel Fishman}
\affiliation{Department of Physics, Technion, Haifa 32000, Israel.}
\affiliation{Department of Physics, University of Maryland, MD 20742, USA.}
\affiliation{Institute for Theoretical Physics, University of California at Santa Barbara, Santa Barbara, CA 93106, USA.}

\date{28 November 2001}

\begin{abstract}
The statistics of energy levels of a rectangular billiard,
that is perturbed by a strong localized potential, are
studied analytically and numerically, when this perturbation
is at the center or at a typical position. Different results
are found for these two types of positions. If the scatterer
is at the center, the symmetry leads to additional
contributions, some of them are related to the angular
dependence of the potential. The limit of the $\delta$-like
scatterer is obtained explicitly. The form factor, that is
the Fourier transform of the energy-energy correlation
function, is calculated analytically, in the framework of
the semiclassical geometrical theory of diffraction, and
numerically. Contributions of classical orbits that are non
diagonal are calculated and are found to be essential. 
\end{abstract}


\maketitle

The distribution of energy levels exhibits a high degree of universality and is a central subject in
the
field of ``Quantum Chaos''~\cite{haakebook,LH89}. For systems that are chaotic in the classical limit
the statistics are of Random Matrix Theory (RMT))~\cite{bohigas84}, while for typical integrable
systems the level distribution satisfies Poissonian statistics~\cite{BT77}. In the semiclassical
regime this universal behavior holds for a wide range in energy. There are also regimes of energy
where spectral correlations related to periodic orbits are important~\cite{berry85,AAA95}. 
In intermediate situations such a high degree of universality in not found. For mixed systems,
where in some parts of phase space the motion is chaotic and in other parts it is regular, the
statistics exhibit some general features~\cite{berry84a,izrailev88}.
Another type of 
intermediate behavior may be found for 
integrable  systems perturbed by singularities of spatial extension that is   
much smaller then
the wavelength of the quantum particle. Examples of relevant  
systems are billiards
with flux lines, sharp corners and
 $\delta$-like interactions~\cite{seba90,bogomolny99E,bogomolny01b}.
Here we report results obtained for a rectangular billiard perturbed by a $\delta$-like
impurity~\cite{long}, known as the \v{S}eba billiard~\cite{seba90}. 
Some of these results can alternatively be concluded from a recent general formulation by Bogomolny
and Giraud~\cite{BG}.

The interest in billiards of various types is primarily theoretical
since it is relatively easy to analyze them analytically and numerically. 
Billiards were studied also experimentally for electrons~\cite{marcus92}, 
microwaves~\cite{stockmann90}
and for laser cooled atoms~\cite{davidson}. We hope that in
the future, perturbations of the type discussed in the present work
will also be introduced experimentally.

Trace formulas that express the quantum density of states in the semiclassical limit as sums
over classical periodic orbits were derived 
for chaotic~\cite{gutz67,gutzwiller} and integrable~\cite{BT-I}
systems.
For perturbations smaller than the wavelength, standard semiclassical theory used in the
derivation of these formulas fails and diffraction
effects
have to be taken into account.
This
can be done in the framework of the Geometrical Theory of
Diffraction (GTD)~\cite{keller62}.
In this approximation, which is valid far from the perturbation,
the Green's function for the system (without the boundary) is given by 
$G(k; {\bf r},{\bf r'}) \simeq G_0 (k; {\bf r},{\bf r'}) + G_0 (k; {\bf r},
{\bf r}_0) D (\theta,\theta')  G_0 (k; {\bf r}_0,{\bf r'})$,
where $\theta$ and $\theta'$ denote the directions of ${\bf r}-{\bf r}_0$
and ${\bf r}_0-{\bf r'}$ respectively and $G_0$ is the free Green's function. The diffraction
constant, $D$, 
describes the scattering from the perturbation.
For the rectangular billiard with a $\delta$-like perturbation, that is subject of the present work, 
{\em only} diffraction effects are responsible for the deviations from the behavior of integrable
systems. Therefore this is an ideal system for the exploration of such effects. Moreover for this
problem
the analytical and numerical calculations are relatively easy. The statistics depend on the location
of the scatterer and on the boundary conditions~\cite{bogomolny01b,berkolaiko01}. This is in contrast to chaotic systems where
the spectral statistics are not affected by such scatterers~\cite{sieber99b}.

The diagonal approximation~\cite{berry85}, where only contributions
from orbits with equal actions are considered 
is extensively used in the field of ``Quantum Chaos''. It 
is not applicable for systems 
with localized perturbations.
 A method to take into account dominant 
non diagonal contributions, in integrable systems, was developed by Bogomolny~\cite{bogomolny00b}
and will be used here.

In this work a rectangular billiard with sides $a_x$ and $a_y$, such that the aspect ratio is
irrational,    
perturbed by a localized scatterer is studied.
The scatterer is represented by a potential of typical size $a$ such that
\begin{equation}
\label{potscale}
U({\bf r}) = \frac{1}{a^2} f \left( \frac{{\bf r}}{a} \right)
\end{equation}
where $f({\bf y})$ is small where ${\bf y}$ is large.

The diffraction constant is the on shell matrix element of the ${\bf T}$
matrix
$ D( \theta',\theta) = \langle {\bf k} | {\bf T}(E) | {\bf q} \rangle $ 
where ${\bf k}$ is the outgoing momentum (in direction $\theta'$) and
${\bf q}$ is the incoming momentum (in the direction $\theta$). 
The energies of the incoming and outgoing waves are equal, that is
$ k=q=\sqrt{E} $
(in units $\hbar=1$ and $m=\frac{1}{2}$ used in this letter).
The Born series cannot be used to compute $D( \theta',\theta)$ 
when $ka \ll 1$ since the free Green's function diverges as $\ln ka$ at short
distances.
A method that is regular when $ka \ll 1$ was introduced by Noyce~\cite{noyce65}.
In this method the scattering in the forward direction is resummed~\cite{noyce65}. 
It leads to a diffraction constant that is a ratio
of series. The series in the numerator and in the denominator are expanded in the number of
scattering events (just like the Born series). Every term in these 
series is then expanded for $ka \ll 1$ (up to terms
of order $k^2a^2$) and both
series are summed (with respect the number of scattering events) to give the angle
dependent diffraction constant~\cite{long}
\begin{widetext}
\begin{equation}
\label{desired}
D(\theta ', \theta)  \simeq  C \left[\rule{0mm}{7mm} 1 +  i \frac{ka}{2} \left\{ \left( e^{i \theta} - e^{i \theta '} \right) M_1 + c.c \right\} 
 -   k^2 a^2 \left(M_0 +  \left\{  \left( e^{2 i \theta} + e^{2 i \theta'}\right)M_2 + c.c. \right\} - \! \! \! \sum_{c,d=-1,+1} \! \! \! M_{cd} e^{i(c\theta' + d\theta)} \right) \right]
\end{equation}
\end{widetext}
where
\begin{equation}
\label{defc}
C \equiv \left( \frac{V(1)}{V_0} + \frac{i}{4}-\frac{1}{2\pi} \left[\gamma + \ln \left( \frac{ka}{2}\right) \right]  + k^2 a^2 \frac{Q(1)}{V_0} \right)^{-1},
\end{equation}
$c.c.$ denotes the complex conjugate, $\gamma$ is Euler's constant and $V_0=\int d^2 y f({\bf y})$.
Also $V(1)$, $Q(1)$, $M_0$, $M_1$, $M_2$ and $M_{cd}$ are constants, independent
of $\theta$, $\theta'$, and  $M_1$, $M_2$ and $M_{cd}$ depend logarithmically on $ka$. These
constants, given by series of integrals, involving the potential (\ref{potscale}), were 
calculated in~\cite{long}.

In the limit $a \rightarrow 0$ and $k$ fixed, a finite diffraction constant is obtained if the
potential is such that,  $\frac{V(1)}{V_0} \sim \frac{1}{2 \pi} [\ln (a/l)+B]$, where
$l$ and $B$ are constants, leading to
\begin{equation}
\label{dfree}
D  \simeq  {2 \pi}\left( \frac{i\pi}{2}- \ln \left( \frac{kl}{2}\right)- \gamma+B \right)^{-1}. 
\end{equation}
It depends on the combination $B-\ln l$ of the two parameters $l$ and $B$. Therefore these are
somewhat arbitrary in the limit $a \rightarrow 0$. 

First we
assume that $a$ is sufficiently small so that (\ref{desired}) can be approximated by (\ref{dfree}),
and because of its slow variation with $k$, it can be replaced by the constant $D$. 
The oscillatory part of the density of states, in the semiclassical limit,
is a sum over contributions of periodic and diffracting orbits. 
Diffracting orbits are orbits which start and return to the scatterer.
For the rectangular billiard with a localized (angle independent) scatterer
at its center
the density of states is~\cite{long}
\begin{eqnarray}
\label{dosc2}
d_{osc} (E) & \! = & \! \sum_p A_p^{(0)} e^{ikl_p} + \sum_{j_1} A_{j_1}^{(1)}  e^{i k l_{j_1}} \nonumber \\ 
& & +   \sum_{j_1,j_2} A_{j_1,j_2}^{(2)}  e^{ik(l_{j_1}+l_{j_2})} \\ & & +   \sum_{j_1,j_2,j_3} A_{j_1,j_2,j_3}^{(3)} e^{ik(l_{j_1}+l_{j_2}+ l_{j_3})}+\cdots +c.c. \nonumber
\end{eqnarray}
where
$A_p^{(0)} = \frac{2 {\cal A}}{\pi \sqrt{8 \pi k l_p}} e^{-i \frac{\pi}{4}}$,
$A_{j_1}^{(1)} = \frac{(-1)^{x_1}\sqrt{l_{j_1}}}{\pi k \sqrt{8 \pi k}} D e^{-i \frac{3}{4} \pi}$,
$A_{j_1,j_2}^{(2)} = (-1)^{x_2} \frac{l_{j_1}+l_{j_2}}{4 \pi^2 k^2 \sqrt{l_{j_1} l_{j_2}}} D^2 e^{-i \frac{3}{2} \pi} $
and
$A_{j_1,j_2,j_3}^{(3)} = (-1)^{x_3} \frac{16 D^3}{3 \pi k (8 \pi k)^{\frac{3}{2}}} 
\frac{l_{j_1}+l_{j_2}+l_{j_3}}{\sqrt{l_{j_1} l_{j_2} l_{j_3}}} e^{-i \frac{\pi}{4}}$, where
$x_1=N_{j_1}+M_{j_1}$, and $x_i=x_{i-1}+N_{j_i}+M_{j_i}$.
The area of the billiard is ${\cal A}$, the length of a periodic orbit is $l_p$ and $l_j$ is the
length of a diffracting
segment $j$ with $N_j$ and $M_j$ reflections from the boundary.
The density of states (\ref{dosc2}) that is expanded to the third order in $D$,
is used to compute the
correlation function
\begin{equation}
\label{correlation}
R_2 (\eta)= \left\langle d_{osc} \left( E-\frac{\eta \Delta}{2} \right) d_{osc} \left( E+\frac{\eta \Delta}{2} \right) \right\rangle \Delta^2
\end{equation}
and its Fourier transform, the form factor
\begin{equation}
\label{formfactor}
K(\tau)=  \int_{-\infty}^{\infty} d \eta R_2 (\eta) e^{2 \pi i \eta \tau},
\end{equation}
where $\Delta$ denotes the mean level spacing.
The brackets denote averaging over an energy scale much larger then $\Delta$
but much smaller then $E$.
If only the contributions from periodic orbits to the density
of states  are taken into account the form factor is given by:
\begin{equation}
\label{nondiagfactor}
\frac{K(\tau)}{2 \pi \Delta} =  \left\langle \sum_{pp'} A_p A_{p'}^{*} e^{{i}(S_p - S_{p'})}
 \delta \left(\frac{2 \pi }{\Delta} \tau - \frac{t_p + t_{p'}}{2} \right) \right\rangle 
\end{equation}
where $S_p$ is the action of the orbit, $t_p$ is its period, and $\tau > 0$ is assumed.
The diagonal approximation can be used to compute the contributions from periodic and once 
diffracting orbits. When there are more then $3$ segments of orbits in the 
exponent of~(\ref{nondiagfactor}), non diagonal contributions are of importance, 
since then one finds saddle manifolds consisting of
different combinations of orbits with almost identical total length so that 
their phase is almost stationary~\cite{bogomolny00b}. 
An example of such a saddle manifold is given by a periodic orbit of length
$l_p=2\sqrt{N_p^2 a_x^2+ M_p^2 a_y^2}$, 
and the pairs of diffracting segments of lengths $l_{j_i} = \sqrt{N_{j_i}^2 a_x^2+ M_{j_i}^2 a_y^2}$
that satisfy
 $N_{j_1}+N_{j_2}=2 N_p$, $M_{j_1}+M_{j_2}=2 M_p$ and $\frac{N_{j_i}}{M{j_i}} \simeq \frac{N_p}{M_p}$. The length difference $l_p-l_{j_1}-l_{j_2}$
is small, of the order of $1/k$. 
Since in billiards the action is $S_j=k l_j$ the action difference is of order unity and these contributions are in phase. 
Other non diagonal contributions of this type can contribute significantly as well.
 The resulting form factor for a scatterer at the center, 
up to order $\tau^3$, 
is found to be ~\cite{long}
\begin{equation}
\label{centerf}
K (\tau) = 1 -\frac{|D|^2}{4} \tau +\frac{1}{8} |D|^4 \tau^2 + \left(\frac{1}{2} |D|^4 - \frac{1}{24} |D|^6 \right) \tau^3.  
\end{equation}
To obtain~(\ref{centerf}) we used the optical theorem, that 
for angle independent scattering is
\begin{equation}
\label{opth}
\Im D =  - \frac{1}{4}|D|^2.
\end{equation}
If the scatterer is 
at a typical location, namely shifted from the center by $(\delta_x a_x,\delta_y a_y)$, with
$\delta_x$, $\delta_y$ and $\delta_x/\delta_y$ all irrational,  the form factor is~\cite{long}
\begin{equation}
\label{typacalf}
K(\tau)= 1-\frac{|D|^2}{4} \tau + \frac{9}{128} |D|^4 \tau^2 + \frac{81}{512} |D|^4 \tau^3-\frac{25}{1536} |D|^6 \tau^3.
\end{equation}
The difference between~(\ref{centerf}) and (\ref{typacalf}) is due to length
degeneracies. When the scatterer is at the center there are four diffracting
segments of identical length, while if it  is moved from the center this degeneracy
is broken. For a quarter of all diffracting segments, for which $N_j$ and $M_j$ are odd, 
the degeneracy is totally lifted. For orbits with even $N_j$ and $M_j$ the location of 
the scatterer does not affect this degeneracy.
For the rest of the segments the degeneracy is only partly lifted. When all
length degeneracies are taken into account one obtains~(\ref{typacalf}). 

\begin{figure}[htb]
\epsfig{file=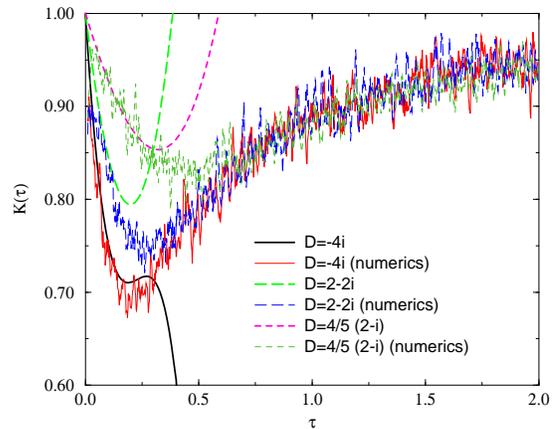,height=6.5cm,width=8cm,angle=0}
\caption{The form factor of a scatterer at a typical position, for some diffraction constants, thin lines, compared to the analytical result~({\protect{\ref{typacalf}}}), heavy lines.\label{noncenter}}
\end{figure}

The form factor can be compared with numerical results obtained for
the case of point interactions, where the eigenvalues are the roots of some function, and therefore
can
be easily found numerically~\cite{seba90}.
The form factor was calculated for several values of $D$ and compared
to the analytical result~(\ref{typacalf}) in Fig.~\ref{noncenter}.
Agreement with~(\ref{typacalf}) is found for short times, as is expected. 

\begin{figure}[htb]
\epsfig{file=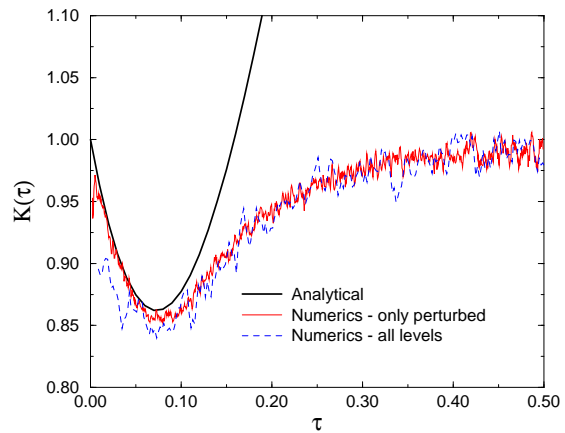,height=6.5cm,width=8cm,angle=0}
\caption{The form factor for the scatterer at the center (${\bf x}_0={\bf 0}$) 
compared to the scaled form factor calculated from perturbed levels~({\protect{\ref{scale}}})
 and the analytical result~({\protect{\ref{centerf}}})\label{centerth}}
\end{figure}

For a scatterer at the center only levels with
wave functions that are symmetric with respect to the $X$ and $Y$ axes
are perturbed by the scatterer.
Since the value of these wave functions for all eigenvalues is the same
($\psi_n ({\bf x}=0)=\frac{2}{\sqrt{\cal A}}$),
 the resulting equation is the same as for the \v{S}eba billiard
with periodic boundary conditions~\cite{bogomolny01b}.
The form factor of the perturbed levels is related to the one of the
full spectrum.
The eigenvalues of the four different symmetry classes of the
rectangle can be assumed to be uncorrelated, leading to
\begin{equation}
\label{scale}
K_{full} (\tau) = \frac{3}{4} + \frac{1}{4} K_{per} (4 \tau).
\end{equation}
The form factor calculated from all levels and the scaled form factor obtained from the perturbed 
levels with the help of~(\ref{scale}) are
compared to the analytical result~(\ref{centerf}), for $D=-4i$, in Fig.~\ref{centerth}.
It is clear that (\ref{scale}) is valid. For very small times the full form factor deviates from its
expected value since there are not enough orbits
that contribute (the calculation is not semiclassical enough). 

In order to study the effect of the angle dependence of the diffraction on the form factor it was 
calculated to the order  $\tau^2$ for the diffraction constant (\ref{desired}).  
For a scatterer  at a typical location the form factor is found to be~\cite{long}
\begin{equation}   
\label{typicalendf}
K (\tau) = 1 - \frac{1}{4} |C|^2 \tau + \frac{9}{128} |C|^4 \tau^2+...~~~.
\end{equation}
This form factor is similar to~(\ref{typacalf}). Since $C$ of (\ref{defc}) satisfies the optical
theorem~(\ref{opth}), 
this form factor can also be obtained from an angle independent potential,
with the diffraction constant $C$.
If the  scatterer is at the center~\cite{long}
\begin{equation}
\label{totalshort}
K(\tau) = 1 -\frac{1}{4} |C|^2 \tau + \frac{1}{8} |C|^4 C' \tau^2+....
\end{equation}
with $C' \equiv 1 - 2 k^2 a^2 M_0$, where $ M_0$ is related to integrals over the potential
(\ref{potscale}).
It resembles the form factor~(\ref{centerf}) that was obtained for angle 
independent scattering. The modification is of the order $k^2a^2$ and typically cannot 
change the sign of the expansion coefficients.

The condition for the applicability of the approximations
used in this letter is
$a \ll \lambda \ll a_x,~a_y$, where $\lambda$ is the wavelength of the particles .
 Up to corrections of order $(ka)^3$, the form factor (\ref{typicalendf}) reduces 
to (\ref{typacalf}) in the order $\tau^2$. 
Therefore the angle dependence plays no role up to this order.
If the scatterer is at the center the situation is somewhat different
as can be seen comparing (\ref{totalshort}) with (\ref{centerf}). There is a correction $C'$
resulting of
the angular dependence of $D(\theta',\theta)$ given by (\ref{desired}). It
is a
consequence of the increased number of length degeneracies of the diffracting
orbits when
the scatterer is at the center. Since the form factor~(\ref{typicalendf}) describes
essentially angle independent scattering the limit $a \rightarrow 0$
describes correctly the physics of the regime $a < \lambda$. This is so
although the
classical dynamics (in the long time limit) are expected to be chaotic in
nearly all of
phase space and similar to the ones of the Sinai billiard. This robustness improves the
chances for the experimental realization of the results of the present work.
For $ a \gg \lambda$, semiclassical theory works and the system should
behave as a
Sinai billiard, with  level statistics given by Random Matrix Theory (RMT) \cite{bohigas84,berry85} (with deviations, see~\cite{sieber93}).

The spectral statistics found in the present work differ from the ones of the
known
universality classes. It is characterized by the form factor of the type
presented in
Figs. \ref{noncenter}
and \ref{centerth}.
This form factor is equal to 1 at
$\tau=0$,
resulting of the fact that for small $\tau$ the number of classical orbits that
are
scattered is small. The contribution that is first order in $\tau$ originates
from the
combinations of forward diffracting orbits and periodic orbits. These always have the same lengths leading to the contribution $\Im D \tau$. By the
optical theorem (\ref{opth}) it is always negative.  For $\tau
\gg 1$ the
form factor approaches unity because of the discreteness of the spectrum~\cite{berry85}.
This general description should hold for other integrable systems, that  are perturbed similarly.

\begin{acknowledgments}
It is our great pleasure to thank E. Bogomolny and M. Sieber for
inspiring,
stimulating, detailed and informative discussions and for informing us about
their
results prior publication. 
We would like to thank also  M. Aizenman, E. Akkermans and R.E. Prange for critical
and informative discussions. 
This research was supported in part by the US-Israel Binational
Science
Foundation (BSF), by the US National Science Foundation under Grant No.
PHY99-07949,
and by the Minerva Center of Nonlinear Physics of Complex Systems.
\end{acknowledgments}

\typeout{References}


\begin{thebibliography}{99}

\bibitem{haakebook}
{Haake F.},
{\em Quantum Signatures of Chaos},
(Springer, New York, 1991).

\bibitem{LH89}
{\em Proceedings of the 1989 Les-Houches Summer School on ``Chaos and Quantum Physics''},
{Giannoni M. J., Voros A. and Zinn-Justin, eds.},
(Elsevier, Amsterdam, 1991).


\bibitem{bohigas84}
{Bohigas O., Giannoni M. J. and Schmit C.},
{\em Phys. Rev. Lett.}, {\bf 52}, 1 (1984).

\bibitem{BT77}
{Berry M. V. and Tabor M.},
{\em Proc. R. Soc. Lond.},
{\bf 356}, 375 (1977).

\bibitem{berry85}
{Berry M. V.},
{\em Proc. R. Soc. Lond.},
{\bf 400}, 229 (1985).

\bibitem{AAA95}
{Andreev A. V. and Altshuler B. L.},
{\em Phys. Rev. Lett.}, 
{\bf 75}, 902 (1995); 
{Agam O., Altshuler B. L. and Andreev A. V.},
{\em Phys. Rev. Lett.},
{\bf 75}, 4389 (1995); 
{Bogomolny E. B. and Keating J. P.},
{\em Phys. Rev. Lett.},
{\bf 77}, 1472 (1996).


\bibitem{berry84a}
{Berry M. V. and Robnik M.},
{\em J. Phys. A: Math. Gen.},
{\bf 17}, 2413 (1984).

\bibitem{izrailev88}
{Izrailev F. M.},
{\em Phys. Rep.},
{\bf 196}, 299 (1990).


\bibitem{seba90}
{\v{S}eba P.},
{\em Phys. Rev. Lett.},
{\bf 64}, 1855 (1990);
{Shigehara T.},
{\em Phys. Rev. E},
{\bf 50}, 4357 (1994).


\bibitem{bogomolny99E}
{Bogomolny E., Gerland U. and Schmit C.},
{\em Phys. Rev. E},
{\bf 59}, R1315 (1999);
{Bogomolny E., Giraud O. and Schmit C.},
{\em Commun. Math. Phys.},
{\bf 222}, 327 (2001);
{Rahav S. and Fishman S.},
{\em Found. Phys.},  
{\bf 31}, 115 (2001);
{Narevich R., Prange R. E. and Zaitsev O.},
{\em Physica E},
{\bf 9}, 578 (2001).

\bibitem{bogomolny01b}
{Bogomolny E., Gerland U. and Schmit C.},
{\em Phys. Rev. E},
{\bf 63}, 036206 (2001).

\bibitem{long}
{Rahav S. and Fishman S.},
{\em ``Spectral statistics of rectangular billiards with localized perturbations''}, to be published.


\bibitem{BG} {Bogomolny E. and Giraud O.},
{\em `` Semi-classical calculations of the two-point correlation
form factor for diffractive systems''},
preprint, nlin.CD/0110006.


\bibitem{marcus92}
{Marcus C. M., Rimberg A. J., Westervelt R. M., Hopkins P. F. and Gossard A. C.},
{\em Phys. Rev. Lett.},
{\bf 69}, 506 (1992);
{Chang A. M.},
{\em Chaos, Solitons \& Fractals},
{\bf 8}, 1281 (1997).

\bibitem{stockmann90}
{St\"ockmann H. -J. and Stein J.},
{\em Phys. Rev. Lett.},
{\bf 64}, 2215 (1990);
{Sridhar S.},
{\em Phys. Rev. Lett.},
{\bf 67}, 785 (1991);
{Richter A.},
{\em Found. Phys.},
{\bf 31}, 327 (2001).

\bibitem{davidson}
{Friedman N., Kaplan A., Carasso D. and Davidson N.},
{\em Phys. Rev. Lett.},
{\bf 86}, 1518 (2001);
{Milner V., Hanssen J. L., Campbell W.C. and Raizen M. G.},
{\em Phys. Rev. Lett.},
{\bf 86}, 1514 (2001).


\bibitem{gutz67}
{Gutzwiller M. C.},
{\em J. Math. Phys},
{\bf 8}, 1979 (1967),
{\bf 10}, 1004 (1969),
{\bf 11}, 1791 (1970),
{\bf 12}, 343 (1971).

\bibitem{gutzwiller}
{Gutzwiller M. C.},
{\em Chaos in classical and quantum mechanics},
(Springer, New York, 1990).

\bibitem{BT-I} 
{Berry M. V. and Tabor M.},
{\em Proc. R. Soc. Lond.},
{\bf 349}, 101 (1976); {Berry M. V. and Tabor M.},
{\em J. Phys. A: Math. Gen.},
{\bf 10}, 371 (1977).



\bibitem{keller62}
{Keller J. B.},
{\em J. Opt. Soc. Am.},
{\bf 52}, 116 (1962).



\bibitem{berkolaiko01}
{Berkolaiko G., Bogomolny E. and Keating J. P.},
{\em J. Phys. A: Math. Gen.},
{\bf 34}, 335 (2001).


\bibitem{sieber99b}
{Sieber M.},
{\em J. Phys. A: Math. Gen.},
{\bf 32}, 7679 (1999), {\bf 33}, 6263 (2000);
{Bogomolny E., Leboeuf P. and Schmit C.},
{\em Phys. Rev. Lett.},
{\bf 85}, 2486 (2000).


\bibitem{bogomolny00b}
{Bogomolny E.},
{\em Nonlinearity},
{\bf 13}, 947 (2000).


\bibitem{noyce65}
{Noyce H. P.},
{\em Phys. Rev. Lett.},
{\bf 15}, 538 (1965);
{Averbuch P. G.},
{\em J. Phys. A: Math. Gen.},
{\bf 19}, 2325 (1986).

\bibitem{sieber93}
{Sieber M., Smilansky U., Creagh S. C. and Littlejohn R. G.},
{\em J. Phys. A: Math. Gen.},
{\bf 26}, 6217 (1993).

\end{thebibliography}
\end{document}